          [2001/04/25 1.1 (PWD)]
\documentclass[a4paper,twocolumn]{esapub2005} 
\usepackage{times}
\usepackage{natbib}
\usepackage{graphicx}
\def\noprintlabel{}
\def\note #1]{{\bf #1]}}
\def\biblab #1{\ifx\noprintlabel\undefined{\bf [#1]}\fi}
\def\cm{\, {\rm cm}}
\def\s{\, {\rm s}}
\def\muHz{\, \mu{\rm Hz}}
\def\Xs{X_{\rm s}}
\def\Zs{Z_{\rm s}}
\def\Ye{Y_{\rm e}}
\def\dcz{d_{\rm cz}}
\def\Rsun{R_\odot}
\def\fig{.}

\title{Prospects for helio- and asteroseismology}
\author{J{\o}rgen Christensen-Dalsgaard}
\affil{Institut for Fysik og Astronomi, Aarhus Universitet, DK-8000 Aarhus C,
Denmark}

\begin{document}


\maketitle

\begin{abstract}
Major progress has been made in helio- and asteroseismology in recent years.
In helioseismology, much of the activity has been in local helioseismology.
However, the recent revision of solar surface abundances, and the resulting
problems in reconciling solar models with the helioseismic inferences,
have lead to renewed activity in solar modelling and global helioseismology.
Interesting, although perhaps not compelling, evidence has been found for
solar g modes in observations with the GOLF instrument on the SOHO spacecraft.
Extensive asteroseismic results have been obtained from ground-based 
observations as well as from the WIRE and MOST satellites, and much is
expected from the upcoming launch of the CoRoT satellite and, in a few
years, from the Kepler mission.
In parallel, stellar modelling is being extended to take some account
of hydrodynamical effects, while large-scale hydrodynamical calculations
are providing increasingly realistic simulations of these effects.
The outcome of these activities will undoubtedly be a far better understanding
of stellar internal properties and stellar evolution, together with an 
improved insight into the physics of matter under the extreme conditions
found in stars.
\end{abstract}

\section{Introduction}
%

The conference demonstrated the impressive and exciting progress that has
been made over recent years in the field of helio- and asteroseismology,
as well as the challenges that we face and the promising prospects of
overcoming them.
Global helioseismology is a mature field, and progress is unavoidably
slower than in the early stages.
The continuing accumulation of data is essential to study the intriguing
variations with time that have been found in the Sun's internal dynamics
but will only very slowly lead to further significant improvements in the data
on the Sun's average properties. 
Such improvements are likely to come rather from improved techniques for
the analysis of the helioseismic data.
At the same time we are faced with increasing challenges to our understanding
of solar modelling.
Local helioseismology is still under rapid development, with an encouraging
increasing emphasis on the understanding of the diagnostic potential and
relative merits of the different techniques.
Asteroseismology has a long and fruitful history in the study of some classes
of stars, particularly the compact pulsators.
However, the development of instruments and the organization of major 
campaigns over the last few years have led to an explosion of activity
on the observational front, with the efforts towards modelling and data
interpretation attempting to keep up.
In all these areas dramatic progress is expected from new satellite missions,
and other instrumentation, to be launched over the coming few years or
in the planning stage.


In the present paper I admit to a strong bias towards global helioseismology and
asteroseismology (although not entirely to the `dark side', as 
Jesper Schou once accused).
Indeed, it is obvious that asteroseismology is a natural extension of
helioseismology in the attempt to understand the large-scale structure and
dynamics of stellar interiors and their evolution with time;
the restriction to far fewer modes in asteroseismic investigations is
to some extent compensated by the ability to observe stars over
a broad range of parameters.
To complement this biased view,
Toomre (these proceedings), in his introduction to the conference, provides
an overview of the accomplishments and promise of local helioseismology
in the study of the more complex three-dimensional and strongly time-dependent
phenomena in the outer parts of the Sun,
including the aptly named `solar sub-surface weather'.
Also, Birch (these proceedings) provides an excellent review of the 
techniques of local helioseismology.


In the analysis of the helio- and asteroseismic data we should keep in
mind that the ultimate scientific goal is to obtain insight into and
understanding of the many complex phenomena that we see in stars.
As tools in this process, and to emphasize the excitement of the field,
we obviously use images illustrating the results, but such images
cannot be goals in themselves. (And indeed, through being graphically
striking, they might even occasionally mislead our intuition, unless we
have a firm understanding of how the graphical representation relates
to the underlying solar or stellar properties.)
It is very encouraging that the conference showed several examples
of a shift towards a comparison of the observational data with, one
may hope, realistic models of phenomena under investigation.
This is an essential part of relating the observations to the
physics of the stellar interiors.


The conference took place just over a year after George Isaak sadly
passed away.
George played a seminal role in the development of helioseismology,
through the initial observations of global five-minute oscillations and
the establishment of the BiSON network and the BiSON group;
he furthermore strongly encouraged the extension of these studies to
other stars.
Thus it was very appropriate that the section of the conference 
dealing with low-degree seismology from the Sun to the stars
was dedicated to his memory.
He would have greatly enjoyed taking part in the conference, with
`a comment and a question' after many of the talks, no doubt often
exposing weaknesses in the argument or pointing the way forward.

\section{In for the long haul}
\label{sec:longhaul}

%

As pointed out by Howe (these proceedings) we now have helioseismic data
spanning more than 30 years, or 1 1/2 solar (magnetic) cycles.
The BiSON and GONG networks and the instruments on the
SOHO satellite have covered a full 11-year sunspot cycle with data with
a very high duty cycle.
As discussed by Howe,
these results have shown intriguing variations in the properties of the solar
interior and the physics of the modes, in most, but not all, cases clearly
related to the sunspot cycle.
The variation of the average multiplet frequencies,
or the component of frequency splittings even in the azimuthal order $m$,
is generally closely correlated with the surface magnetic field
\citep{Bachma1993, Chapli2001, Howe2002}.
However, in his presentation Verner 
showed that for low-degree modes there is a clear hysteresis between
the frequency shift and various indicators of solar activity
(Chaplin et al., these proceedings),
as has previously been found \citep{Jimene1998, Tripat2001}.
The long time basis in
the results presented by Verner allowed the demonstration of
interesting variations in the relation between frequency shift and activity
between solar cycles.

The solar internal rotation shows more striking variations with time.
The clearest signal are the zonal flows, regions of slightly slower and
faster rotation (Howe, these proceedings).
At low and intermediate latitude a similar pattern in the solar
surface velocity has previously been
identified as solar `torsional oscillations' \citep{Howard1980};
from helioseismic analyses it is now known that these flows extend over
a substantial fraction of the convection zone.
They converge towards the equator as the sunspot cycle progresses and
indeed the location of the flow 
is close to, but shifted slightly towards lower latitudes than, the
sunspot belts, possibly suggesting that the flows lead the activity in
the development of the solar cycle \citep{Howe2006}.
At higher latitudes the corresponding surface feature had not previously
been identified; there the bands of slower and faster rotation move towards
the pole with the progression of the sunspot cycle.
The physical causes for these flows are discussed by Rempel (these proceedings);
the high-latitude component appears to be
dominated by the feedback of the Lorentz force on the differential rotation,
whereas no entirely
satisfactory explanation has been found for the behaviour at lower latitudes.

As discussed by Howe (these proceedings) an oscillation with a period of
1.3 yr has been detected near the base of the convection zone, at the equator.
This was observed consistently in independent analyses during the period
1996 -- 2001 while the behaviour in recent years has been more erratic.
Basu \& Antia \citep{Basu2001},
while seeing a superficially similar behaviour in
their results, argued that it was not significant, and indeed the rather 
erratic nature of the observed variation might reasonably raise concerns
about the physical reality of the 1.3-yr oscillation.
On the other hand, it is not unreasonable that oscillations in a potentially
strongly coupled and nonlinear system display time variations, either 
tied to specific phases in the solar cycle or in a more chaotic fashion.

Given these time variations, strongly related to the solar cycle and of 
a somewhat irregular nature, it is obviously important to continue 
detailed helioseismic analyses on a timescale covering several solar cycles,
and in a form that ensures homogeneous datasets.
For the ground-based observations this requires maintaining both the
BiSON and GONG networks; it is important to emphasize that these two networks
complement each other, the BiSON network providing data on the lowest-degree
modes that are superior to those obtained by the spatially resolved
observations.
A further important incentive for extending the BiSON observations is the 
push to lower frequencies in the p-mode range. 
These modes have lifetimes
of several years and hence their frequencies can be determined with extremely
high precision; 
thus they are potentially very valuable for the determination of the
core structure and rotation.
The space-based observations clearly have a limited lifetime; however the
GOLF observations on SOHO should be continued as long as possible, and
a reasonable overlap must be ensured between the MDI observations on SOHO 
and the HMI observations on SDO, to ensure an adequate intercalibration
of these two datasets.

\section{Improved data for global helioseismology}
\label{sec:data}

%
Global helioseismology has been remarkably successful in uncovering 
the detailed structure and rotation of the solar interior.
Yet, unavoidably, these inferences have revealed a need for even
finer details, to allow a deeper understanding of the underlying
phenomena.
Results on the structure of the inner core are ambiguous, at a level which
may have some effects on the use of solar models to constrain neutrino
properties.
The determination of the rotation of the core is rather uncertain, leaving
room for a small core rotating substantially faster (or perhaps more
slowly) than the bulk of the radiative interior.
The important issues of penetration and overshoot at the base of the
convection zone, not least the potential variation with latitude, have
only been addressed in fairly crude terms.
Also, substantial uncertainty remains concerning the detailed properties
of the so-called tachocline, i.e., the transition region 
between the latitudinally varying rotation in the convection zone and
the near-constant rotation in the radiative interior,
and its location relative to the base of the convectively unstable envelope.

Needless to say, resolving these issues requires better data;
in particular, not only random, but also potential systematic, errors
must be reduced.
In principle, random errors can be reduced through extending the
observations \citep{Libbre1992},
although a significant improvement at the present stage will obviously
require perhaps unrealistically long time series (at least on the 
natural timescale of the present author).
Also, the variation of the frequencies over the solar cycle makes it
difficult to combine data over such long periods.
Indeed, the study of these time variations is in itself of great interest
and clearly requires a determination of accurate frequencies from 
time series of limited duration.
Thus it is of obvious interest to develop techniques that allow 
reliable frequency determinations, or perhaps more direct inferences
about the solar interior, from the existing observations.

The analysis of helioseismic observations is complicated by the 
inability, owing to the restriction of the observations to
one hemisphere of the Sun, to separate fully the contributions from modes of
different degree $l$ and azimuthal order $m$ in the spatial analysis.
Thus the analysis of the time series for a given target $(l_0, m_0)$
must take into account the cross-talk from modes of other $l$ and $m$.
In earlier analyses this has been done in a somewhat incomplete way
\citep{Anders1990, Schou1992, Schou1999}.
Also, damping and the stochastic nature of the excitation of the modes
must be taken into account, leading, for example, to an asymmetry in
the peaks in the power spectra.
If not treated properly, such effects may cause systematic errors in
the frequencies which could lead to serious problems in the inferences
of the solar internal properties.

\begin{figure}
\centering
\includegraphics[width=1.0\linewidth]{\fig/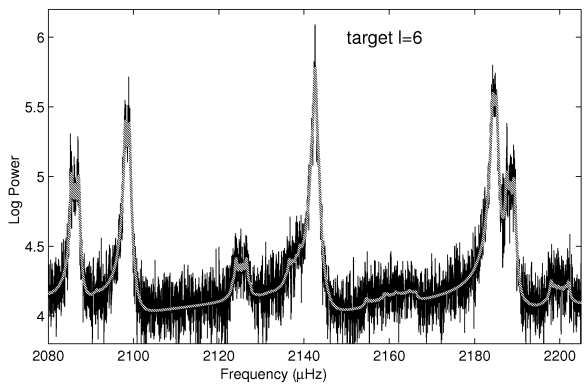}
\caption{
Model of a one-year $m$-averaged power spectrum from the MDI instrument on
the SOHO spacecraft, for spherical harmonic degree $l = 6$ and radial
order $n = 12$.
The thin line shows the data and the heavier grey line shows the fit.
Adapted from \citep{Voront2005}.
\label{fig:vorfit}
}
\end{figure}

Substantial progress towards resolving these problems
have been made by Jefferies, Vorontsov \& Giebink (these proceedings),
who report results from a major project to develop techniques for 
analysis of global helioseismic data which, as far as possible,
take the known properties of the modes into account.
The cross-talk is taken fully into account, using techniques
developed in \cite{Voront2005}.
Also, the physics of the excitation and damping is described in some
detail, resulting in a good representation of the spectral line profiles.
Finally, the solar structure is directly included in the fit, through
a phase function corresponding to the inferred solar seismic model.
An example of the resulting fit to the observed power spectrum is
illustrated in Figure~\ref{fig:vorfit}.
The results obtained by Jefferies et al.\ show a marked reduction in
the frequency residuals resulting from the fit, compared with the
previously available frequencies, and provide very interesting
inferences of the sound-speed correction relative to a solar model
and the buoyancy frequency, even in the inner solar core.

The analysis by Jefferies et al.\ considered
mean multiplet frequencies and concentrated on the structure of the
solar interior.
A major challenge will be to extend the analysis to individual 
frequencies, in this way determining the rotational splitting and
possible structural departures from spherical symmetry.
Another important issue, particularly for the study of the solar
core, is the consistent combination of different datasets, particularly
the spatially resolved data from GONG or MDI with the more detailed and
accurate low-degree data from the unresolved observations with BiSON or GOLF.
Such tasks must have very high priority.
There seems little doubt that the full development of these techniques
will very considerably strengthen our ability to infer the interior
structure and dynamics of the Sun, including the determination of,
or constraints on, possible time variations.


High-degree modes are particularly sensitive to instrumental effects and,
for ground-based observations, effects of the Earth's atmosphere.
Thus, despite their strong potential as diagnostics of the outer
parts of the convection zone \cite{Rabell2000},
they have so far seen little use in global helioseismology.
It is encouraging that Rabello-Soares et al.\ (these proceedings)
report progress on the understanding and elimination of instrumental
effects for the MDI instrument on SOHO.
If combined with the analysis techniques developed by Jefferies et al.
we may hope finally to obtain reliable data on high-degree modes for
use in inverse analyses.
This will be particularly important for the study of the thermodynamics
of the hydrogen and helium ionization zones.

\section{Solar g modes?}

%
Solar g modes have been of the nature of a `holy grail' of helioseismology
ever since the early claims to the detection of a 160-minute oscillation
\citep{Brooke1976, Severn1976, Scherr1979}.
The reason for this interest is obvious: g modes have their largest
amplitudes in the solar core and hence potentially provide much more
detailed information about the structure and rotation of the deep solar
interior.
This is perhaps most important in the case of rotation, where no 
information is obtained from radial modes which penetrate most deeply,
and where the low-degree nonradial p modes have very limited sensitivity
in the inner core.
Consequently, there have been extensive efforts and further claims
to the detection of g modes \citep{Delach1983, Frohli1984}.
There seems little doubt that the early claims were premature;
a stringent upper limit to the amplitude of possible g modes,
well below the earlier apparent detections, was obtained in \cite{Appour2000}.

Theoretical predictions of the g-mode amplitudes are extremely uncertain
but tend to support this non-detection. Gough \cite{Gough1985}
and Kumar et al.\ \cite{Kumar1996} predicted amplitudes around
$0.1 \cm \s^{-1}$ and $0.01 \cm \s^{-1}$, respectively, at a
frequency of $200 \muHz$, with a rapid decrease of amplitude at
lower frequencies.
These estimates were based on extending the stochastic excitation by
near-surface convection, which is quite successful for the p modes
(cf.\ Section~\ref{sec:excitation}), to lower frequencies. 
Excitation of gravity waves also likely occur at the base of the
convection zone (see Talon, these proceedings),
although most efficiently at the longer periods corresponding to
the characteristic convective time scales in the lower parts of the
convection zone.
In any case, these estimates are probably sufficiently uncertain that
they cannot be taken as strict requirements for an observational search for
g modes.


Due to their potential diagnostic importance, g-mode detection
has been an important goal of the GOLF instrument on the SOHO
spacecraft \cite{Gabrie1995}.
The analysis of the GOLF data has in fact led to a renewed interest
in the search for solar modes at low frequencies, including g modes.
Turck-Chi\`eze et al.\ \cite{Turck2004} found tentative evidence
for multiplet structure at very low amplitudes, at frequencies
corresponding to low-order p and g modes.
In these proceedings, Garc\'{\i}a et al.\ present an analysis that instead
attempts to identify the expected overall structure of high-order g modes.
For such modes the periods are approximately uniformly spaced, at
a spacing that depends on the degree.
By detecting this structure in the observations the presence of
g modes can be confirmed, and the properties of the modes, including
the average period spacing and possibly individual frequencies, can
be determined.
Such analyses have been extensively used in earlier searches for,
and claimed detections of,
solar g modes; however, the analysis by Garc\'{\i}a et al.\ probably
represents the first time that they are applied to the GOLF data.

\begin{figure}
\centering
\includegraphics[width=1.0\linewidth]{\fig/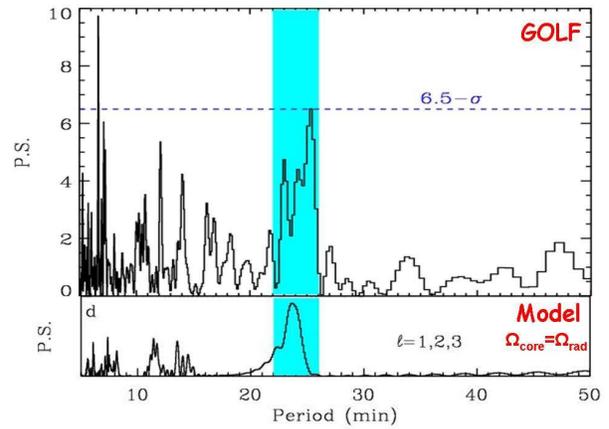}
\caption{
Power spectrum of power-spectrum density in the g-mode frequency range.
The lower panel is for a solar model, while the upper panel is based on
10 years of observations with the GOLF instrument on the SOHO spacecraft.
The shaded area indicates the location of the peak corresponding to dipolar
g modes equally spaced in period.
(From Garc\'{\i}a et al., these proceedings.)
\label{fig:gmodes}
}
\end{figure}

Garc\'{\i}a et al.\ analyse a 10-yr time series of GOLF data; they 
remove a solar background from the power spectrum and carry out a
Fourier transform of the resulting spectrum, as a function of period,
to identify possible periodic structures, as expected from asymptotic theory.
To ensure that asymptotic theory is valid, while the expected 
amplitudes are not too small, the analysis is restricted to the
cyclic frequency interval between 25 and $140 \muHz$.
Analysis of corresponding artificial data based on a solar model
shows a strong signal only at the period spacing corresponding to $l = 1$.
As illustrated in Figure~\ref{fig:gmodes} the observed frequencies
show a similar response at approximately the same period spacing as for
the model.


Given the extensive analysis required to obtain this result,
and the potentially complex sources of noise, including solar `noise',
it is evidently important to test its significance of the result.
Based on a Monte Carlo simulation with a large number of
noise realizations Garc\'{\i}a et al.\ find that
the observed $l = 1$ peak is significant at a level of 99.5 \%.
This is encouraging; however, further tests are probably required
using different assumptions about the character of the noise background.
As a related issue, it is found that the observed spectrum of the
spectrum can be reproduced in the presence of background noise only
if the lifetimes of the modes are relatively short, of the order of
a few months. 
This is substantially below the expected damping times of these modes;
however, it could reflect phase variations of the modes on this
timescale, and caused by variations in the internal properties of the Sun.
The possible 1.3-yr oscillation in the rotation rate indicates that
such variations may be present.

From model fits to the observed data Garc\'{\i}a et al.\ conclude that
the core rotation of the Sun may exceed the surface value by as much
as a factor five. 
This would obviously be a very interesting result but probably requires
further confirmation, as well as tests of whether it is consistent with
existing constraints from p-mode splittings on the rotation of the
deep solar interior \citep{Chapli1999}.

Unfortunately, the analysis does not provide estimates of the
oscillation amplitudes, and hence comparisons cannot be made with earlier upper
bounds on the g-mode amplitudes or with the theoretical predictions.
Such estimates can presumably be obtained from further simulations,
perhaps with different assumptions about the frequency dependence
of the amplitudes; 
they are urgently needed to strengthen the case for the reality of this signal.
However, it is certainly encouraging that evidence for g modes appears
to be present in the GOLF data, and further analyses and simulations
in this direction will be of great interest.

\section{Near-surface effects}
\label{sec:nearsurface}

%
The near-surface regions of the Sun are obviously highly complex, with a strong
interplay between thermal structure, flows and magnetic fields.
The richness and challenges of this region are reviewed by 
Erd\'elyi (these proceedings).
Thus the ability to make detailed investigations of these regions
by local helioseismology, reviewed by Toomre (these proceedings) and discussed
by a number of papers in the present volume, is extremely interesting.
At the photospheric level, and presumably below, magnetic effects are
strongly localized to active regions where the magnetic and gas pressure
are comparable. 
Particularly important is the critical surface where the sound speed equals
the Alfv\'en speed and where consequently mode conversion occurs.
The detailed interaction
between the magnetic field and the acoustic waves depends critically on
the inclination angle of the field and the direction of propagation of the
waves \citep{Schunk2006}.
Jefferies et al.\ (these proceedings) have found that such effects allow
the propagation of waves into the solar chromosphere at frequencies below
the acoustical cut-off frequency (around 5 mHz) in the non-magnetic part
of the atmosphere.
Such `magneto-acoustical portals' are associated with the small-scale 
strong magnetic features that are ubiquitous on the solar surface.
Jefferies et al.\ demonstrate that the wave energy transmitted through
these regions at relatively low frequency may play a substantial role
in the heating of the chromosphere.
Furthermore, this effect likely has a significant influence on the damping,
and possibly the frequencies, of global modes of oscillation.


A striking parallel to these solar effects, albeit on a far larger scale,
is discussed by Kurtz et al.\ (these proceedings), in connection with the 
rapidly oscillating Ap stars.
Here the properties of the global modes are dominated by the large-scale
magnetic field in these stars.
Strikingly, the strong separation of elements with altitude, and across
the stellar surface, allows detailed inferences to be made of the
variation of the oscillation eigenfunctions with height in the atmosphere,
thus providing potentially very stringent tests of the understanding
of the physics of the modes under the influence of the magnetic field.


The interaction between the magnetic fields and the acoustic waves
poses particular problems for the interpretation of results of
local helioseismology in the vicinity of active regions.
Lindsey \& Braun \cite{Lindse2005} identified an acoustical
`shower-glass effect', resulting from the distortion of the acoustic waves
by subphotospheric magnetic fields,
which impairs the analysis, at least in the case of helioseismic
holography (see also Lindsey, these proceedings).
These effects are further discussed by Schunker \& Cally (these proceedings),
while Zhao (these proceedings) notes that similar effects are present
in time-distance analyses, although possibly less severe.
Related problems concern the interpretation of results of local helioseismology
in terms of direct magnetic, or induced thermal, effects.
These issues must be understood before a fully reliable 
interpretation, in physical terms, can be made of the results of
local helioseismology.


The goal of helioseismology, global as well as local, is obviously to
resolve the properties of the solar interior.
However, the physical properties of the observed modes unavoidably place 
constraints on the resolution that can be achieved.
Unresolved aspects of the near-surface effects must 
somehow be suppressed in the analysis of the observations to infer properties
deeper in the solar interior.
In the case of structure inversions in global helioseismology it is
well known that the dominant difference between the observed and the model
frequencies is generally caused by unresolved effects near the surface
(e.g., \citep{Christ1988}).
From the frequency dependence of these effects they must predominantly be
localized to the uppermost parts of the solar interior and the solar atmosphere,
with likely culprits being convective effects where the stratification
is strongly superadiabatic or effects of the mode damping and excitation.
In particular, modes of relatively low frequency, which are reflected below this
region, suffer minimal effects.
These effects are unlikely to be fully resolved in the radial direction
and hence must be suppressed in the inverse analysis;
in the case of global helioseismology this is largely possible owing
to the availability of modes over a broad range of frequency and degree.
The frequency effects can be regarded as corresponding to a phase shift
of the waves when they are reflected near the surface;
as such they should be present also in the properties of waves as studied
in local helioseismology, including time-distance analysis;
it is interesting that this has recently been confirmed \citep{Braun2006}.


These near-surface effects pose serious problems in the analysis of 
data on solar-like oscillations in distant stars, where only data on
low-degree modes are available.
Even here the effects, insofar as the investigations of stellar cores are
concerned, can to some extent be suppressed through a suitable 
combination of the data \citep{Roxbur2003, Oti2005, Roxbur2005}.
However, a great deal of information, particularly about the global properties
of the star such as its radius, is lost in the process.
Based on the solar experience, one may expect that lower-frequency modes
are less affected and hence provide largely unbiased diagnostics;
however, these modes generally have lower amplitudes and hence are
challenging to observe.

Evidently, the ultimate goal is to understand these effects 
and hence use them as additional diagnostics.
Promising progress in this regard has been made on the basis of 
hydrodynamical simulations of convection in the solar
\citep{Rosent1999, Robins2003} and stellar \citep{Straka2006} case.
The results presented by Stein et al.\ (these proceedings)
have taken such simulations to a new level of detail.
On this basis it should be possible to investigate directly the effects
of convection on the properties of the modes,
by analysing simulations extending over a sufficiently long period that the
modes excited in the simulation has time to reach a suitable equilibrium.
This could obviously be done for a range of stellar parameters, thus
possibly allowing a reliable calibration of simpler formulations for
more general use.
Also, additional effects such as magnetic fields could be included in the
simulations.

\section{Compositional challenges}

Helioseismology has provided what must be regarded as a relatively accurate
determination of properties of the solar internal structure,
particularly the sound speed.
Although the inferences most often are made with reference to solar models,
leading to a determination of the differences in, e.g., sound speed between
the Sun and the model, the final sound-speed profile resulting from 
applying that correction to the model is largely insensitive to the reference
model \citep{Basu2000}.
There remain issues of resolution and error correlation which mean that 
testing of a specific model is best carried out by using that model as
reference in an inverse analysis to determine the differences with the
solar structure, as discussed in \citep{Christ2005}.
However, in general helioseismology has been remarkably successful in
determining the solar internal sound speed.

\begin{figure}
\centering
\includegraphics[width=1.0\linewidth]{\fig/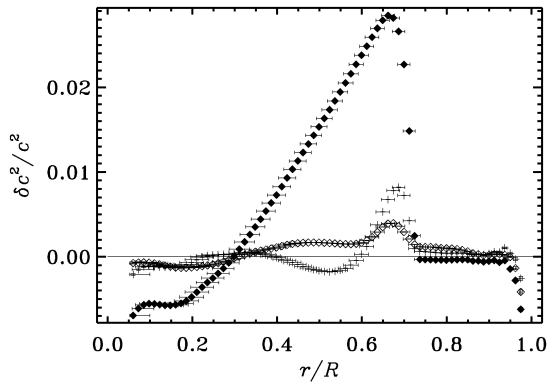}
\caption{
Relative differences in squared sound speed, in the sense (Sun) - (model),
against fractional distance $r/R$ to the solar centre.
The open diamonds are for Model S \citep{Christ1996}.
The filled diamonds are for a model with the revised composition presented
by Grevesse, while the crosses are for a model where in addition the neon
abundance has been increased by 0.5 dex.
The (barely visible) vertical bars show $1-\sigma$ errors, whereas the
horizontal bars give a measure of resolution in the underlying inversion.
(Houdek et al., in preparation.)
\label{fig:solsound}
}
\end{figure}

The success of a solar model then depends on how well it reproduces that
helioseismic determination.
In parallel with the development of helioseismology calculations of solar
evolution were improved, leading to models that were in reasonable agreement
with helioseismic inferences, although still differing from them by much more
than the formal errors in the inferences (see Figure~\ref{fig:solsound});
a typical example is `Model S' \citep{Christ1996} which has been extensively
used in helioseismic analyses.
Although the improvements in the computations were inspired by helioseismology
no explicit parameter adjustments were made to match the helioseismic results;
the input to the calculation was obtained from the, at the time, best possible
knowledge about the physics of the solar interior and the other observables,
such as the ratio $\Zs/\Xs$ between the present surface heavy-element and
hydrogen abundances.
By adjusting parameters or aspects of the internal physics
models in even closer agreement with the
helioseismic sound speed can be constructed (e.g., \citep{Turck2001}).


Recently, a major revision of the determinations of solar surface abundances
has taken place \citep{Asplun2005a, Asplun2005b, Greves2005}, 
as summarized by Nicolas Grevesse at the conference.
The improvements in the abundance analyses involve using three-dimensional
and time-dependent hydrodynamical simulations with a realistic treatment
of radiative transfer, as well as abandoning the assumption of
local thermodynamic equilibrium 
in the treatment of the formation of some spectral lines.
A strong indication of the improvement in the line-formation modelling
resulting from the use of the hydrodynamical models is evident from the
fact that the line profiles can be modelled without the use of 
{\it ad hoc\/} micro and macro turbulence.
Also, strong support for these new results come from the fact that,
unlike earlier determinations,
consistent abundances are obtained for a given element from different
spectral lines. 
The most dramatic change is in the abundance of oxygen which is reduced
by a factor of 0.54 in the new determination, relative to
the values used until recently in solar modelling \citep{Anders1989};
also, the carbon abundance is reduced by a factor of 0.68.
As a result $\Zs/\Xs$ is changed from the value 0.0245 used, for example, in
Model S to $\Zs/\Xs = 0.0165$.

The heavy-element abundances predominantly affect solar structure through
the opacity; in fact, oxygen makes a dominant contribution to the opacity
just beneath the convection zone.
Thus it is not surprising that the change in composition leads to a change
in solar structure, as illustrated in terms of sound speed in 
Figure~\ref{fig:solsound};
clearly, the composition change has caused a strong deterioration in the
agreement between the model and the Sun.
Also, the initial hydrogen abundance required to calibrate the model is
changed, leading to a decrease in the helium abundance $\Ye$ in the
convective envelope from 0.245 in Model S to 0.221, and hence to a present
surface heavy-element abundance $\Zs = 0.0125$ 
(compared with 0.0181 in Model S).
Finally, the depth $\dcz$ of the convective envelope changes from
$0.289 \Rsun$ ($\Rsun$ being the solar radius) in Model S to
$0.271 \Rsun$.
As discussed, for example, in \citep{Bahcal2006}
the values for $\Ye$ and $\dcz$ obtained for models like Model S
were consistent with helioseismic inference while the new values are not.


This renewed discrepancy between solar models and the helioseismically
determined solar structure is a serious problem for solar and stellar modelling.
Guzik (these proceedings) discusses it in more detail, including the many
attempts that have been made to solve it.
There is clearly a need for independent confirmation of the analysis by
Asplund et al.;
in fact, the result has been questioned on the basis of possible problems
with the solar atmospheric temperature profile used \citep{Ayres2006}.
On the assumption that the revised abundances are correct,
it is 
evident that, since the problem arises from the change in opacities resulting
from the change in composition, it can be solved by an intrinsic opacity
modification that compensates for the change in the heavy-element abundance.
The magnitude of the required opacity change is probably larger than the
quoted precision of the opacity calculations, although perhaps not larger
than the real uncertainty in the computed opacities.
Such a change is hardly satisfying without a physical justification;
however Kurucz (private communication) has noted that
the very many weak lines that are neglected in current opacity calculations
might add up to a contribution of the required size, thus strongly
motivating consideration of such effects.
Opacity may also be increased by increasing abundances of other elements
making substantial contributions to the opacity.
Interesting examples are neon and argon, whose solar abundances are very 
uncertain, since they have no spectral lines formed in the solar photosphere.
It has been shown that, for example, an increase in the neon abundance by
0.5 dex (i.e., a factor 3.16) comes close to restoring the agreement 
between the model and the Sun seen for Model S \citep{Antia2005, Bahcal2005};
an example of such a model is also illustrated in Figure~\ref{fig:solsound}.
However, although there may be evidence for a such a higher abundance of
neon in solar-neighbourhood stars \cite{Drake2005} the applicability of
these results to the Sun has been questioned (e.g., \citep{Schmel2005}).
A further test of the abundance of neon in the solar convection zone
can in principle be obtained from helioseismic constraints on
the effect on the equation of state,
particularly the adiabatic compressibility \citep{Lin2005}.
An initial analysis indicates that the required enhanced neon abundance
is inconsistent with the helioseismic results \citep{Antia2006}.

Guzik (these proceedings) considers several other potential solutions
to the problem posed by the new abundances, including energy transport
against the temperature gradient by gravity waves, hence corresponding to
a higher effective opacity
(interestingly, gravity waves have also been invoked to account for the
transport of angular momentum; see Talon, these proceedings).
She concludes that the previous reasonable (though not perfect) agreement
between the solar models and the Sun might have arisen from a fortuitous
cancellation of errors, and that the new discrepancies are a good motivation
to reconsider the computation of solar models, including also normally 
neglected hydrodynamical effects.
This clearly represents an exciting, if difficult, challenge.

\section{Physics of stellar interiors}
\label{sec:physics}


Opacity is not the only aspect of the `microphysics' used in stellar modelling
that might be in question. 
The issue of electron screening of nuclear reactions in stellar interiors
has seen vigorous discussion over an extended period
(e.g., \citep{Shaviv1996}, \citep{Brugge1997}, \citep{Tsytov2000},
\citep{Bahcal2002}).
Most calculations of stellar evolution use the classical description
of the screening enhancement developed by Salpeter \citep{Salpet1954},
based on an equilibrium treatment of the average charge around the
reacting nuclei. 
This description may, however, reasonably be questioned on the basis of the
relevant timescales and the moderate number of particles typically involved
in the screening.
This has motivated investigations of the interactions through 
molecular-dynamics calculations, particularly by Shaviv and Shaviv
\citep{Shaviv2000, Shaviv2004}.
Given the complexity of such calculations, it is very encouraging that
Mussack et al.\ (these proceedings) have embarked on similar but entirely
independent calculations.
The results are eagerly awaited.


Although the microphysics is complex and challenging in stellar
evolution calculations,
even more serious challenges are caused by potential hydrodynamical
and magneto-hydrodynamical phenomena in stellar interiors.
Convection is an obvious difficulty; as discussed by
Stein et al.\ (these proceedings) an apparently realistic description is
possible of the near-surface convection, although at large computational
expense.
Also, Toomre (these proceedings) presents results of large-scale simulations
of convection and rotation in the solar convection zone, including also
magnetic effects, although not yet giving a fully adequate modelling of,
for example, the solar internal differential rotation.
Such detailed simulations of the deeper interiors of stars are hampered
by the vast range of timescales that need to be considered, between the
overall thermal timescale and the dynamical timescales.
Thus no fully realistic simulations have been made, for example, of
convective penetration beneath the solar convection zone.

Observationally, it is clear that non-standard phenomena must be 
considered in stellar modelling.
The nearly uniform rotation of the solar radiative interior requires
mechanisms for the transport of angular momentum towards the surface,
to reduce internal rotation from the assumed initial rapid rate in
the early Sun.
Also, in many cases the observed surface abundances of stars require
mixing beyond the convectively unstable regions, hence indicating 
some kind of hydrodynamical instability.
Palacios et al.\ (these proceedings) provide an overview of the required 
non-standard extensions of stellar modelling.

A dominant source of dynamical effects in many stars is rotation.
It is relatively straightforward to include the spherically symmetric
component of the centrifugal force in stellar modelling, although
this requires a model for the evolution of the internal rotation rate with age. 
A much more difficult issue is the meridional circulation and the 
instabilities which are likely caused by rotation and which lead
to transport of chemical elements and angular momentum.
Zahn \citep{Zahn1992} developed an approximate treatment of these
processes which has been refined and extensively applied.
Interestingly, the resulting transport of angular momentum is
insufficient to account for the solar internal rotation
(see Palacios, these proceedings).

Schatzman \citep{Schatz1993} proposed that gravity waves excited
by penetration from the solar convective envelope might be efficient
in transporting angular momentum in the radiative interior,
thus accounting for the internal solar rotation.
As discussed by Talon (these proceedings) further refinements, including
differential filtering of the waves in a layer just below the convection
zone, are required to make this a potentially viable mechanism. 
The internal rotation rate resulting from such calculations is quite
similar to the helioseismically inferred behaviour.
Alternatively, magnetic fields may play an important role in coupling
the interior to the convection zone.
Gough \& McIntyre \citep{Gough1998} pointed out that even a weak
primordial field would be sufficient and furthermore discussed how this
could account for the localized transition of rotation in the solar
tachocline (see Toomre, these proceedings).
Alternatively, the field may be generated by dynamo action through
differential rotation in the radiative interior \citep{Spruit2002};
such a model also results in near uniform rotation of the radiative interior,
as observed in the Sun \citep{Eggenb2005}.

The processes discussed here should of course be included in a complete
modelling of stellar structure and evolution.
Mathis et al.\ (these proceedings) describe an ambitious programme to
develop a stellar evolution code making this possible,
with an initial application to a 1.5 $M_\odot$ star.
Development and verification of such a code is evidently a major task
which is urgently needed, given the extensive asteroseismic results 
expected from upcoming missions.
The observational tests of the results will require analysis of a broad
range of observables in different types of stars.
The solar internal structure and rotation obviously provide strong 
constraints.
Additional constraints are available from the detailed distribution of
element and isotope abundances observed in different stars, reflecting
the internal nuclear reactions and the mixing processes bringing the
products of these reactions to the surface, as well as the destruction
of lithium and possibly beryllium through nuclear burning.
Also, the detailed asteroseismic results expected from CoRoT and Kepler,
and eventually from network projects such as SONG
(see Section~\ref{sec:prospects}),
should provide strong constraints on the modelling.

It should be kept in mind that the current and planned level of modelling,
although relatively complex, still relies on highly simplified descriptions
of the magneto-hydrodynamical processes in stars.
Thus the modelling should be complemented by investigations aiming at
improving our understanding of these processes, to improve their description
in the stellar evolution codes.
This will undoubtedly involve substantial numerical simulations,
the results of which,
while perhaps not directly applicable to stellar conditions, can at least
be extrapolated to such conditions.


\section{Mode excitation and damping}
\label{sec:excitation}

%
Two different mechanisms are available to excite a given mode:
self-excitation where the mode is linearly unstable, with an amplitude
that consequently grows as a result of the conversion of thermal energy
into mechanical energy;
and external forcing of an intrinsically damped mode.
In the former case a critical region of the star acts as a heat engine,
with the appropriate phasing of heating and compression during the pulsation
cycle.
In order for the effect to lead to overall instability that region 
must be located at a suitable depth within the star; 
consequently, stars pulsating because of heat-engine excitation tend
to be located in relatively well-defined regions of the HR diagram
where that condition is satisfied.
A typical example is the classical Cepheid instability strip where the
excitation is caused by opacity features associated with the second
ionization of helium.
The external forcing of stable modes can take place through tidal
interaction in binary systems.
In single stars, which are the focus here, the only mechanism 
that has been considered in any detail is forcing by near-surface convection;
it should be recalled, however, that excitation of gravity waves 
through convective penetration beneath convective envelopes may play
a role for stellar structure and evolution (cf.\ Section~\ref{sec:physics}).

Dupret et al.\ (these proceedings) review the properties of modes
excited through the heat-engine mechanism.
In relatively hot stars, where convection at most makes a minor contribution
to energy transport in the relevant regions of the star, 
the critical aspect of the energy transport leading to excitation is
the behaviour of opacity, and one therefore often talks about the
`$\kappa$ mechanism'.
Excitation is typically associated with rapid variations, or `bumps',
in the opacity near a given temperature;
instability arises for modes such that the oscillation period matches
the thermal timescale of that part of the star which is outside the
region where local excitation takes place \citep{Cox1974}.
Each characteristic bump in the opacity is generally associated with
one or more classes of pulsating stars;
longer-period modes are evidently preferred when the bump occurs deeper
in the star, i.e., in cooler stars.
As discussed below,
examples of this behaviour are found in the region of the $\beta$ Cephei
and slowly pulsating B stars, as well as amongst the pulsating
subdwarf B stars.

In modelling cooler pulsators, the treatment of the interaction between
convection and pulsations is a major challenge.
This concerns both the perturbation to the convective flux, which affects
the heating, and the perturbation to the turbulent pressure, which has a 
direct dynamical effect on the pulsations.
As discussed by Dupret et al.\ convective effects are likely responsible
for the return to stability at the cool edge of the Cepheid instability strip.
This is confirmed by a number of independent calculations, using rather
different physical descriptions of how pulsations affect convection.
However, the detailed cause of the damping differs between the different 
formulations; in some calculations (e.g., \citep{Houdek2000})
the perturbation to the turbulent pressure dominates the damping, while
in others (e.g., \citep{Dupret2005}) damping is dominated by
the convective flux. 
It is evidently essential that these issues be clarified through detailed
comparisons between the different calculations.
Also, it is important to ensure that the convection is treated in a
manner that is consistent between the calculation of the equilibrium model
and the pulsations.

In the $\gamma$ Dor stars, beyond the cool side of the instability strip,
the mode excitation is apparently caused by a heat-engine mechanism, 
but controlled by the effect of convection.
In the critical layer at the base of the convective envelope the convective
timescale is much longer than the pulsation period, and hence convection
does not react to the pulsations. 
This `frozen convective flux' leads to heating just beneath the convection
zone in the appropriate phase of the pulsations, and hence to instability
\citep{Guzik2000}.
Dupret et al.\ note that calculations with a detailed treatment of the
reaction of convection confirm this simple picture. 
It should be noticed that this is an example of heat-engine excitation that
cannot be regarded as an opacity mechanism.

Heat-engine excitation demonstrates instability of a mode but says nothing
about the limiting amplitude.
Except for large-amplitude pulsators, where the excitation mechanism
apparently saturates, there is no definite understanding of the
amplitude-limiting mechanism or the resulting distribution of mode
amplitudes.
In a detailed analysis Nowakowski \citep{Nowako2005} showed that
resonant mode interaction between unstable and stable modes was
not able to account for the observed amplitudes, as had previously been
suggested.
Thus a reliable theory for the mode selection in heat-engine pulsators
remains an important challenge to the study of such stars.

%

The stochastic excitation of solar-like oscillations is reviewed
by Houdek (these proceedings).
In this case the amplitudes are determined by the balance between
the energy input from convection and the damping rate, leading 
to a definite prediction of the pulsation amplitudes.
The calculation of the damping rate evidently requires treatment
of convective effects which, as already mentioned, is uncertain.
In the solar case, however, accurate observational determinations of
the damping rates have been obtained from the measured line widths
of the modes.
As illustrated by Houdek, computed damping rates, based on the
convection theory of Gough as further developed by Balmforth 
\citep{Gough1977, Balmfo1992}, can be adjusted to be in reasonable
agreement with these observations.
This gives some hope that the description may be extended to other stars.
In fact, the computed damping rates for $\alpha$ Cen A and B are similar
to those inferred from observations (and to the solar values).
On the other hand, for the giant star $\xi$ Hydrae the computed lifetime
\citep{Houdek2002} is considerably longer than the lifetime
inferred, albeit somewhat indirectly, from observations \citep{Stello2006}.
The observational determination of lifetimes will be greatly improved by
the expected long-duration observations from, e.g., the CoRoT and Kepler
missions (see Section~\ref{sec:prospects}), which will allow a direct
determination of the line widths.
However, it is obvious that improvements are also required in the computation
of the damping rates.

\begin{figure}
\centering
\includegraphics[width=1.0\linewidth]{\fig/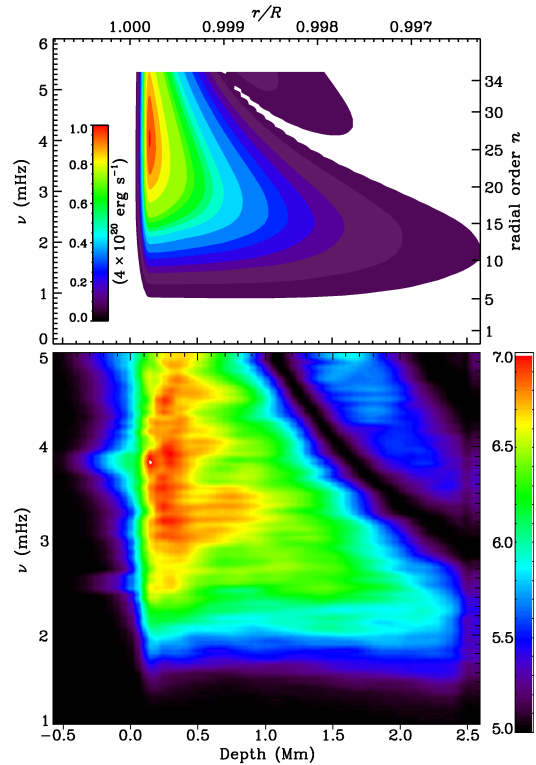}
\caption{
Energy input in stochastic excitation of solar acoustic modes
as a function of depth below the photosphere and cyclic frequency $\nu$.
The upper panel shows results for a simple mixing-length model of
convection, on a linear scale, whereas the lower panel,
on a logarithmic scale, is based on
hydrodynamical simulations of near-surface convection \citep{Stein2001}.
(Adapted from Houdek, these proceedings.)
\label{fig:excitation}
}
\end{figure}


The energy input comes partly from dynamical effects through the
Reynolds stresses, and partly from thermodynamical effects through
fluctuations in the buoyancy force (often referred to as entropy fluctuations).
Computation of these contributions obviously requires a model for
the temporal and spatial spectra of convection in the equilibrium model,
as discussed by Houdek.
Since this involves higher-order moments of the turbulent field,
appropriate closure models are also required;
an interesting recent development is the use of closure models that
explicitly take into account the asymmetry between broad upwellings and narrow
downdrafts in stellar convection \citep{Belkac2006}.
Further insight can be obtained from comparing the simple models,
typically based on some version of mixing-length theory, with the energy
input computed from detailed hydrodynamical models of convection
\citep{Stein2001}, as illustrated in Figure~\ref{fig:excitation}.
It is evident that the general behaviour is similar in the two cases,
although excitation in the simulations extends to considerably
greater depth than in the simple model.

In the solar case, the computed amplitudes are generally
in reasonable agreement with observations;
however, if the computed damping rates are used the predicted amplitudes
are substantially larger than the observed values at low frequencies,
where in fact the computed damping rates are too small.
Using instead the observed damping rates, the predicted and observed
amplitudes are in good agreement \citep{Chapli2005}.
For other stars the situation is less clear, as summarized by Houdek.
In particular, it appears that the predicted amplitudes are substantially
higher than observed for stars hotter than the Sun, such as Procyon.
Interestingly, Houdek notes from hydrodynamical simulations that there
may be some cancellation, so far not included in the calculations,
between the contributions from Reynolds stress and buoyancy to the
excitation; this can perhaps explain at least partly why the computed
values are too high.
In the case of $\xi$ Hydrae it is striking that the predicted amplitude
is in excellent agreement with observations, despite the large discrepancy
in the damping rates.

An issue closely related to mode excitation, also considered by Dupret et al.
and by Houdek,
is the relation between observable properties of the oscillations as
seen in intensity or radial velocity.
This is important for the use of different observations of a given mode
to identify the degree of the mode.
The observed properties obviously depend on the response of the atmosphere
to the oscillations.
At frequencies well below the acoustical cut-off frequency a quasi-static
treatment of the oscillations in the atmosphere is probably adequate
\citep{Dupret2001}, 
while at higher frequencies, such as are observed in the Sun and solar-like
pulsators, a fully time-dependent calculation may be required
\citep{Christ1983}.
Also, the treatment of the convective flux is obviously important
in cooler stars;
an interesting analysis of the effect on mode identification in
the $\delta$ Scuti star FG Vir was carried out in \citep{Daszyn2005}.
Additional complications in the mode identification arise in rapidly
rotating stars, where a given mode may have contributions from more
than one spherical-harmonic degree \citep{Daszyn2002}.

It is evident that the near-surface and atmospheric regions are extremely
complex in stars with convective envelopes, yet crucial for the understanding
and observations of oscillations in such stars;
this also, as discussed in Section~\ref{sec:nearsurface}, applies to
the interpretation of the oscillation frequencies.
Although the theoretical understanding of these regions is improving,
there is no
doubt that major advances will result from the hydrodynamical simulations.
The new simulations of solar near-surface convection reported by
Stein et al.\ (these proceedings) are a gold-mine for the further development
of studies of excitation and other properties of stellar oscillations.
Oscillations are spontaneously excited in the simulation and the duration,
65 hours of solar time, may be sufficient to allow a determination of
the damping rate of the modes and an understanding of the dominant factors
controlling the excitation and damping \citep{Georgo2003}.
With similar simulations for selected examples of other stars we can hope
to be able to calibrate simpler models and in this way obtain reliable
calculations of mode properties for a wide range of stars.
When compared with the upcoming large observing campaigns this provides
excellent prospects for an improved understanding of the properties
of the oscillations and, one may hope, observational constraints on the
properties of convection.

\section{Progress on asteroseismology}

%
The remarkable recent development in asteroseismology covers most of the
HR diagram and a broad range of stellar evolutionary stages.
Developments in ground-based techniques for radial-velocity observations
have contributed greatly to this, as have substantial programmes for
coordinated observation of selected objects.
Also, as reported by Jaymie Matthews at the conference
the MOST satellite \citep{Walker2003}
has produced striking results for a substantial number of stars,
although perhaps not quite reaching the expected level of sensitivity
towards solar-like oscillations.

Bedding \& Kjeldsen (these proceedings) discuss the stochastically
excited solar-like oscillations 
which are likely found in all stars with extensive outer convection zones.
It is striking that such oscillations, with the characteristic
Lorentzian profiles, have likely been detected in red supergiants such 
as Betelgeuze ($\alpha$ Ori), from amateur observations recorded
by the American Association of Variable Star Observers
(see also \citep{Kiss2006}).
Solar-like oscillations in red giants are well established;
a remarkable set of data from the MOST satellite has been obtained
for the star $\epsilon$ Ophi (Barban et al., these proceedings).
Interestingly, Hekker et al.\ (these proceedings) find from 
radial-velocity observations of the same star that the observed modes
are likely non-radial; 
this raises problems for the theoretical understanding of oscillations
in red giants, according to which such modes are 
damped substantially more strongly than are radial modes \citep{Dziemb2001}.

Closer to the main sequence detailed frequency spectra have been obtained
for a substantial number of stars.
An interesting case is $\mu$ Ara which also hosts a planetary system and,
like many other planetary hosts, has a relatively high heavy-element
abundance $Z$.
As discussed by Vauclair (these proceedings) the asteroseismic analysis in
principle allows to distinguish between models with a globally high $Z$ and
models where just the convection zone is so enriched;
in practice the current data, although extensive, are still not quite 
sufficiently detailed to allow a decision on this issue
(see also \citep{Bazot2005}).
In an analysis of an extended series of data from the WIRE satellite
Fletcher et al.\ (these proceedings;
see also \citep{Fletch2006}) were able to determine directly the
average mode lifetime and rotational splitting for $\alpha$ Cen A.
Potentially very interesting, although controversial,
results on $\eta$ Boo have been reported
based on MOST observations \cite{Guenth2005};
these indicate a sequence of radial modes
extending from the previously observed modes towards modes of low order.
Such modes would have very valuable diagnostic potential;
however, their detection at relatively high amplitudes is 
contrary to our understanding of stochastic excitation and entirely 
unlike the properties of the spectra observed in other stars with
solar-like oscillations.
Thus some scepticism towards this claimed detection is probably warranted.


In contrast to the solar-like oscillators, the heat-engine pulsators
(see also Section \ref{sec:excitation}) occupy relatively well-defined
instability regions in the HR diagram, determined by the physical
mechanisms responsible for the excitation.
Also, these stars show pulsations at larger amplitudes than the solar-like
pulsators near the main sequence (however, smaller-amplitude pulsations,
below the current observational threshold, are likely also present;
e.g., \citep{Breger2005}).
Matthews presented extensive results from MOST for a broad range of stars,
including $\delta$ Scuti and $\beta$ Cephei stars;
very interesting results have been obtained for the rapidly oscillating
Ap star HD~1217, showing additional `anomalous' frequencies affected by
magnetic effects.
Aerts (these proceedings) discusses oscillations in hot main-sequence stars.
Detailed analyses of $\beta$ Cephei stars have become possible thanks
to very extended sets of coordinated observations over the last few years.
An interesting case is $\nu$ Eri where standard models do not find
instability of all the observed modes in models that fit the observed
frequencies \citep{Aussel2004, Pamyat2004};
the instability is likely caused by enhancements of the abundances
of iron-group elements in the driving region, through settling and
radiative levitation, as is also invoked for the subdwarf B pulsators
(see below).
Significant asteroseismic results can be obtained from even a small
number of modes, if these are properly identified;
thus Mazumdar et al.\ \citep{Mazumd2006} obtained stringent constraints,
including a determination of the core overshoot,
for the $\beta$ Cephei star $\beta$ CMa from just three modes of which
two were shown to be undergoing an avoided crossing.
Slowly pulsating B (SPB) stars have been discovered in substantial numbers
but their very long periods (of the order of a day or more) make a
detailed analysis very difficult and time consuming.
Also, while the identification of the observed frequencies
with high-order g modes makes them potentially interesting diagnostics,
the very dense spectrum of theoretical frequencies and the 
generally rapid rotation (comparable to the oscillation frequencies)
of the stars complicate the interpretation of the observations.
An extensive set of frequencies in the star HD~163830
has been obtained from the 
MOST observations, but a definite identification of the degree of the modes,
let alone identification with modes in stellar models, has not yet been
possible \citep{Aerts2006}.

Fontaine (these proceedings) discusses the pulsating subdwarf B (sdB) stars,
a recently identified class of variable stars.
These are helium-burning stars at the extreme blue end of the horizontal branch 
who have lost almost all of the hydrogen-rich envelope, 
possibly as a result of binary evolution.
The generally extensive sets of observed frequencies provide an excellent
potential for studying the properties of these highly evolved stars
and hence constraining the earlier evolutionary phases.
Two classes of pulsating sdB stars have been found: short-period variables
pulsating in p modes and, at slightly lower effective temperature,
long-period variables pulsating in high-order g modes.
The excitation of the modes can be understood in terms of heat-engine
driving in the opacity bump associated with iron-group elements, but only
if the abundances of these elements are enhanced by radiative levitation
\citep{Fontai2003}.
The required enhancement is established on a timescale short 
compared with the evolutionary timescale of these stars \citep{Fontai2006}.

%


\section{Asteroseismic diagnostics}

%
Reliable extraction of information about stellar properties from the
observed oscillation frequencies requires a good understanding of the
relation between those properties and the properties of the oscillation
frequencies.
This greatly aids the combination of frequencies in ways that 
isolate measures of specific aspects of the stellar interiors,
with an understanding of how the observational uncertainties 
give rise to uncertainties in these measures.
An obvious example is the extensive use of the large and small frequency
separations in asteroseismic diagnostics, based on solar-like oscillations,
to determine global parameters of the stars.

Evidently the goals of asteroseismology, particularly given the upcoming
observational projects, go much beyond such basic global parameters.
Consequently, we need to understand how the finer details of the observed
frequencies are related to more subtle aspects of the stellar interior.
An important example, discussed by Houdek \& Gough (these proceedings),
is the effects on the frequencies of relatively sharp features, 
denoted `acoustical glitches' by Houdek \& Gough.
Such features include the effect on the adiabatic compressibility of
the ionization of helium, and the abrupt change in the temperature and
hence sound-speed gradient at the base of a convective envelope.
Houdek \& Gough develop analytical expressions for the influence
of these effects on the second difference of frequencies of low-degree
modes of like degree.
The expressions contain parameters that, for example, provide information
about the helium content of the convective envelope and its depth.
They provide excellent fits to second differences of computed frequencies 
for a solar model.
The quality of the fit, and the physical basis for the parametrization,
is undoubtedly superior to previous examples of similar analyses
\citep{Montei1998, Basu2004, Montei2005}.
Thus it seems likely that the parametrization by Houdek \& Gough will
provide a more reliable determination of these properties in stellar interiors,
once adequate data are available.


Rigorous results in the theory of stellar oscillations are rare, for
general stellar models.
In most cases insights into the properties of the modes are based on
asymptotic analysis or numerical calculations.
Thus the announcement by Takata \citep{Takata2005} of an identity
strictly satisfied between the perturbation quantities of oscillations
of degree $l = 1$ is a major advance in the theory of stellar oscillations.
Such dipolar oscillations are special in having a geometry where
the centre of mass of each perturbed shell in the star is 
displaced, while obviously the centre of mass of the star as a whole
remains stationary during the oscillation.
This property forms the basis for the identity.
As a result of the identity, the equations of adiabatic oscillations
can be reduced from the usual fourth-order system to a second-order system.
In these proceedings Takata further develop this theory in a 
conceptually simpler form, which can be directly compared with the
classical asymptotic description in the Cowling approximation of
neglecting the perturbation to the gravitational potential.
This will undoubtedly improve our understanding of the diagnostic 
potential of these modes;
it is worth recalling that amongst acoustic modes the dipolar modes,
together with the radial modes, are those that penetrate most deeply
into the stellar core.

\begin{figure}
\centering
\includegraphics[width=1.0\linewidth]{\fig/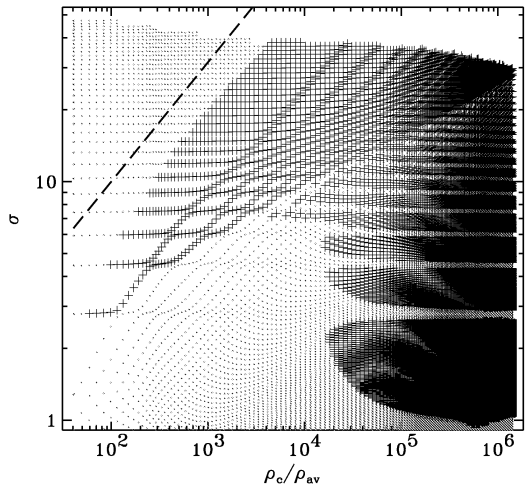}
\caption{
Computed dimensionless frequencies $\sigma$ for dipolar modes
for a $1 M_\odot$ evolution sequence against central condensation;
here $\sigma = (R^3/GM)^{1/2} \omega$,
$\omega$ being the angular frequency, $R$ and $M$ the stellar radius and mass
and $G$ the gravitational constant, and central condensation is
measured as the ratio between the central and average densities
$\rho_{\rm c}$ and $\rho_{\rm av}$.
Small dots mark modes that are labelled correctly by the usual
scheme \citep{Scufla1974, Osaki1975}, whereas crosses mark modes where this
scheme fails.
The dashed line is defined by $\sigma^2 = \rho_{\rm c}/\rho_{\rm av}$.
(Christensen-Dalsgaard \& Takata, in preparation.)
\label{fig:dipolar}
}
\end{figure}

The reduction of the order of the equations to two also allows a simple
procedure for defining the radial order of the dipolar modes.
The commonly used procedure, essentially based on assuming the
Cowling approximation, fails for models of high central condensation
(e.g., \citep{Guenth1991}, \citep{Christ1994}).
In contrast, applying a similar procedure to the variables satisfying
the exact second-order equations derived by Takata allows a robust
assignment of order for any model.
Such a secure labelling further helps the understanding of how the 
modes change as the star evolves and allows tests to ensure that no
modes have been missed in computations of stellar oscillation spectra.
As an example, Figure~\ref{fig:dipolar} illustrates where the previously
used procedure fails in an evolution sequence for a $1 M_\odot$ model.

\section{Challenges and prospects}
\label{sec:prospects}

%
It is obvious that great progress has been made towards an improved
understanding of the interior of the Sun and other stars, through 
extensive observations and modelling.
Equally obviously, many challenges remain.
The treatment of rotation remains a critical issue, despite the progress
on including rotational effects in stellar modelling.
For the many stars rotating substantially faster than the Sun the treatment
of the effect of rotation on the oscillation frequencies requires going
beyond the first-order expansion typically used in helioseismology.
An expansion up to third order has been developed \citep{Soufi1998}
but has so far, perhaps because of its complexity, seen relatively limited use.
Also, even this expansion is inadequate for many of the stars for
which asteroseismic investigations are carried out;
here a fully two-dimensional calculation is required to match the
observational precision \citep{Lignie2006, Reese2006}.
Similarly, a consistent fully two-dimensional calculation of stellar
structure and evolution will be needed;
some progress has been made in this direction
(e.g., \citep{Deupre1990}, \citep{Roxbur2004}, \citep{Jackso2005}),
although much work obviously remains before this can be combined with
the treatment of hydrodynamical effects discussed in Section~\ref{sec:physics}.
The development of this required next level of stellar modelling
will undoubtedly
take place through a combination of idealized, if highly complex,
hydrodynamical simulations addressing specific aspects of the physics
and comprehensive attempts to model the full stellar evolution
with somewhat simplified representation of the physics.

%
The observational prospects for helio- and asteroseismology are bright.
As argued in Section~\ref{sec:longhaul} the BiSON and GONG networks
should be kept operating to ensure coherent data covering several
solar cycles.
With the extension of SOHO operations through 2009 a reasonable overlap
should be ensured with the observations from 
the Solar Dynamics Observatory (SDO).
The launch of SDO, expected in late 2008, will result in a dramatic increase
in data on solar waves and oscillations, with Helioseismic and Magnetic
Imager (HMI) yielding full-disk Doppler-velocity data with a 
spatial resolution of 1.5 arc seconds and a time resolution of 
less than one minute.
Unlike the MDI instrument on SOHO such high-resolution data will be available
continuously throughout the mission.
This will revolutionize the possibilities for following the variations in
solar subsurface features in time with local helioseismology,
provided that the required development of efficient and reliable analysis
techniques, with the ability to keep up with the data flow, takes place.
Also, the data have great potential for global helioseismology,
including the analysis of the near-surface region with high-degree modes,
once techniques to extract reliable and accurate frequencies of such modes
have been established (see Section \ref{sec:data}).

In the longer term the Solar Orbiter mission is planned to observe the Sun
from a distance of as little as 0.22 AU and, equally importantly,
from an orbit inclined at up to $35^\circ$ 
from the ecliptic \citep{Marsch2005}.
This will for the first time permit detailed observations of the solar polar
regions,
including the possibility of using local helioseismology to measure
flows at high latitude, such as the near-surface meridional circulation
which may be crucial for the action of the solar dynamo 
(Rempel, these proceedings).
Also, since the spacecraft is orbiting the Sun it will allow time-distance
seismology with a very long baseline, and hence providing information about
deep layers in the Sun, including the tachocline region.
However, the large distance from Earth and the resulting low data-transmission
rate will be a limiting factor for such investigations.
It is hoped to launch Solar Orbiter in 2015, although this will depend on 
the evolution of the ESA Science Programme in a situation of serious
financial constraints.

The DynaMICS plans outlined by Turck-Chi\`eze et al.\ (these proceedings)
intend to combine several instruments for the study of the solar interior
and atmosphere with the goal to understand, and possibly predict, the 
long-term trends in solar variability.
The proposed instruments include the GOLFNG, a further development of the
GOLF instrument on SOHO; a prototype of the instrument is being constructed
and will be tested within the next year in Tenerife.
The goal is to carry out more detailed studies of the deep solar interior,
ideally through the observation of g modes, and providing information about 
the dynamics and possible magnetic fields of the radiative interior.

The WIRE and MOST missions have demonstrated the potential for asteroseismology
from space, based on intensity observations.
This potential will be further exploited with the launch of the CoRoT
satellite towards the end of 2006.
In particular, the CoRoT long runs, lasting 5 months,
will yield asteroseismic data of unprecedented quality for a variety of
stars and hence allow much more detailed investigations of such stars
than previously possible.
Additional data for a large number of stars will be obtained from the
Kepler mission, scheduled to be launched by NASA in late 2008 
\citep{Basri2005}.
Kepler will observe a field with an area of around 100 square degrees in
the constellations Cygnus and Lyra continuously for at least 4 years.
The mission is designed to investigate extrasolar planetary systems,
and in particular search for Earth analogues, by means of transit detections.
To optimize the chances of detection of transits more than $100\,000$
stars will be observed at a cadence of 30 minutes.
However, observations of 512 stars will be carried out at a cadence of
1 minute, thus, for example, allowing the detection and study of solar-like 
oscillations.
The field contains thousands of stars for which it is estimated that such
oscillations can be investigated.
The selection of stars can be changed every three months; however, it is
expected that some stars will be observed throughout the mission, to
improve the quality of the data and search for possible frequency variations
associated with cyclic stellar activity.
In addition, the observations with 30-minute cadence will obviously be 
sufficient for the study of solar-like oscillations in giant stars,
as well as many other types of stellar variability.
The Kepler dataset will undoubtedly become a unique resource for the study
of almost any kind of small-amplitude stellar variability.

Beyond Kepler, there are plans to submit a proposal for the
PLATO project (PLAnetary Transits and Oscillations of stars) 
for the expected ESA call for new mission proposals.
PLATO will observe a field 10 times as large as Kepler, with the goal
of detecting and studying large numbers of planetary systems as well
as carrying out asteroseismic investigations of a very large number
of relatively bright stars.


Experience from helioseismology has shown the importance of obtaining
data of the highest possible precision and extending over very extended
periods.
Indeed, it is obvious from the analysis reported by
Houdek \& Gough (these proceedings) that a frequency accuracy of 
the order of $0.1 \muHz$ will be required, for example, for the
analysis of the signatures of acoustic glitches.
Also, it is highly desirable to extend the data towards lower frequencies;
here the lifetime of the modes is generally longer, leading to more
precise frequency determinations, the data are more sensitive to
the effect of helium ionization, and the frequencies are less affected
by the near-surface problems.
Improving the frequency precision depends on increasing the signal-to-noise
ratio in the observations.
This evidently requires minimizing the instrumental noise;
however, an ultimate limit to the noise level is provided by the
stellar `noise', i.e., the effects of phenomena in the stellar atmosphere
other than the oscillations.
Already Harvey \citep{Harvey1988} pointed out that such stellar noise
is far more important, relative to the oscillation amplitudes,
in intensity than in velocity observations.
The results from the VIRGO and GOLF instruments on SOHO have clearly
demonstrated this, the signal-to-noise ratio being higher by almost
two orders of magnitude in velocity than in intensity.
Thus it is highly desirable to observe the oscillations in velocity.

For such observations there is not much to be gained from observing from
space.
However, the obvious challenge is to obtain the required continuous
observations extending over long periods.
With existing observational facilities this require coordination between
different time-allocation committees;
also, the required very stable spectrographs tend to be on large telescopes
where obtaining more than a week's observing time is extremely difficult.
Thus dedicated facilities are needed.
Nearly continuous observations from a single site can be obtained from
Antarctica.
This is the basis for the SIAMOIS project
(for Sismom\'etre Interf\'erentiel A Mesurer les Oscillations 
des Int\'erieurs Stellaires) to be set up at Dome C, a near-optimal
site for astronomical observations (Mosser, these proceedings).
Here a duty cycle of over 90 \% is expected during three months in mid-winter.

An alternative is to follow the lead of the helioseismic networks, such
as BiSON and GONG, and establish an asteroseismic network of telescopes
dedicated to Doppler-velocity observations of stellar oscillations.
To investigate this possibility we have established the SONG project
(for Stellar Oscillation Network Group).%
\footnote{see {\tt http://astro.phys.au.dk/SONG}}
The project is currently in an initial phase of conceptual design
for such a network,
including preliminary designs of the telescopes and spectrographs.
The Doppler-velocity observations will use iodine cells as reference;
it is estimated that with one-meter-class telescopes solar-like oscillations
can be studied with good signal-to-noise ratio in stars brighter than
6th magnitude.
Ideally 8 nodes will be required, 4 in each hemisphere, to ensure good
coverage of the entire sky.
The intention is to make the nodes fully robotic.
The nodes will be established at existing observatory sites, to ensure
that the required infrastructure is available and that limited maintenance
can be obtained on site.

It should be noticed that the SONG project and the
space-based photometric missions are fully complementary.
Photometric observations can be made simultaneously of a large number of
stars, providing excellent statistics of stellar properties.
However, only through Doppler-velocity observations dedicated to carefully
selected stars and carried out over long periods can we achieve the
full potential of asteroseismology to probe the details of the physics of
stellar interiors. 
The combination of these techniques, and the parallel development of
stellar modelling, will surely take our knowledge of stellar structure
and evolution, and of the extreme physical processes that control
stellar interiors, to new levels of understanding in the coming years
and decades.

\section*{Acknowledgements}

I am grateful to the organizers for giving me the challenge of providing an
overview of a remarkably rich and exciting conference.
I thank many authors for providing me with their papers
for these proceedings before publication, R. A. Garc\'{\i}a, G. Houdek and 
S.~Vorontsov for help with figures,
and A. Weiss for making explicit to me the point
that the new solar abundances are
a problem for solar modelling, not for helioseismology.


\begin{thebibliography}{}

\bibitem[Aerts {\it et~al.\/}(2006)]{Aerts2006}
\biblab{Aerts2006}
Aerts C., De Cat P., Kuschnig R., et al., 2006,
{\it ApJ}, {\bf 642}, L165 

\bibitem[Anders \& Grevesse(1989)]{Anders1989}
\biblab{Anders1989}
Anders E. \& Grevesse N., 1989,
{\it Geochim. Cosmochim. Acta}, {\bf 53}, 197 

\bibitem[Anderson {\it et~al.\/}(1990)]{Anders1990}
\biblab{Anders1990}
Anderson E. R., Duvall T. L. \& Jefferies S. M., 1990,
{\it ApJ}, {\bf 364}, 699 

\bibitem[Antia \& Basu(2005)]{Antia2005}
\biblab{Antia2005}
Antia H. M. \& Basu S., 2005,
{\it ApJ}, {\bf 620}, L129 

\bibitem[Antia \& Basu(2006)]{Antia2006}
\biblab{Antia2006}
Antia H. M. \& Basu S., 2006,
{\it ApJ}, {\bf 644}, 1292 

\bibitem[Appourchaux {\it et~al.\/}(2000)]{Appour2000}
\biblab{Appour2000}
Appourchaux T., Fr\"ohlich C., Andersen B., et al., 2000,
{\it ApJ}, {\bf 538}, 401 

\bibitem[Asplund(2005)]{Asplun2005a}
\biblab{Asplun2005a}
Asplund M., 2005,
{\it Annu. Rev. Astron. Astrophys.}, {\bf 43}, 481 

\bibitem[Asplund {\it et~al.\/}(2005)]{Asplun2005b}
\biblab{Asplun2005b}
Asplund M., Grevesse N. \& Sauval A. J., 2005,
in {\it Cosmic Abundances as Records of Stellar Evolution and Nucleosynthesis},
eds T. G. Barnes III \& F. N. Bash, ASP Conf. Ser. {\bf 336},
p. 25 

\bibitem[Ausseloos {\it et~al.\/}(2004)]{Aussel2004}
\biblab{Aussel2004}
Ausseloos M., Scuflaire R., Thoul A. \& Aerts C., 2004,
{\it MNRAS}, {\bf 355}, 352 

\bibitem[Ayres {\it et~al.\/}(2006)]{Ayres2006}
\biblab{Ayres2006}
Ayres T. R., Plymate C. \& Keller C. U., 2006,
{\it ApJS}, {\bf 165}, 618 

\bibitem[Bachmann \& Brown(1993)]{Bachma1993}
\biblab{Bachma1993}
Bachmann K. T. \& Brown T. M., 1993,
{\it ApJ}, {\bf 411}, L45 

\bibitem[Bahcall {\it et~al.\/}(2002)]{Bahcal2002}
\biblab{Bahcal2002}
Bahcall J. N., Brown L. S., Gruzinov A. \& Sawyer~R.~F., 2002,
{\it A\&A}, {\bf 383}, 291 

\bibitem[Bahcall {\it et~al.\/}(2005)]{Bahcal2005}
\biblab{Bahcal2005}
Bahcall J. N., Basu S. \& Serenelli A. M., 2005,
{\it ApJ}, {\bf 631}, 1281 

\bibitem[Bahcall {\it et~al.\/}(2006)]{Bahcal2006}
\biblab{Bahcal2006}
Bahcall J. N., Serenelli A. M. \& Basu S., 2006,
{\it ApJS}, {\bf 165}, 400 

\bibitem[Balmforth(1992)]{Balmfo1992}
\biblab{Balmfo1992}
Balmforth N. J., 1992,
{\it MNRAS}, {\bf 255}, 603 

\bibitem[Basri {\it et~al.\/}(2005)]{Basri2005}
\biblab{Basri2005}
Basri G., Borucki W. J. \& Koch D., 2005,
{\it New Astronomy Rev.}, {\bf 49}, 478 

\bibitem[Basu {\it et~al.\/}(2000)]{Basu2000}
\biblab{Basu2000}
Basu S., Pinsonneault M. H. \& Bahcall J. N., 2000,
{\it ApJ}, {\bf 529}, 1084 

\bibitem[Basu \& Antia(2001)]{Basu2001}
\biblab{Basu2001}
Basu S. \& Antia H. M., 2001,
{\it MNRAS}, {\bf 324}, 498 

\bibitem[Basu {\it et~al.\/}(2004)]{Basu2004}
\biblab{Basu2004}
Basu S., Mazumdar A., Antia H. M. \& Demarque~P., 2004,
{\it MNRAS}, {\bf 350}, 277 

\bibitem[Bazot {\it et~al.\/}(2005)]{Bazot2005}
\biblab{Bazot2005}
Bazot M., Vauclair S., Bouchy F. \& Santos N. C., 2005,
{\it A\&A}, {\bf 440}, 615 

\bibitem[Belkacem {\it et~al.\/}(2006)]{Belkac2006}
\biblab{Belkac2006}
Belkacem K., Samadi R., Goupil M. J., Kupka~F. \& Baudin F., 2006,
{\it A\&A}, in the press {\tt [astro-ph/0607570 v3]}.

\bibitem[Br\"uggen \& Gough(1997)]{Brugge1997}
\biblab{Brugge1997}
Br\"uggen M. \& Gough D. O., 1997,
{\it ApJ}, {\bf 488}, 867 

\bibitem[Braun \& Birch(2006)]{Braun2006}
\biblab{Braun2006}
Braun D. C. \& Birch A. C., 2006,
{\it ApJ}, {\bf 647}, L187 

\bibitem[Breger {\it et~al.\/}(2005)]{Breger2005}
\biblab{Breger2005}
Breger M., Lenz P., Antoci V., et al., 2005,
{\it A\&A}, {\bf 435}, 955 

\bibitem[Brookes \& Isaak(1976)]{Brooke1976}
\biblab{Brooke1976}
Brookes J. R., Isaak G. R. \& van der Raay H. B., 1976,
{\it Nature}, {\bf 259}, 92 

\bibitem[Chaplin {\it et~al.\/}(1999)]{Chapli1999}
\biblab{Chapli1999}
Chaplin W. J., Christensen-Dalsgaard J., Elsworth~Y., et al., 1999,
{\it MNRAS}, {\bf 308}, 405 

\bibitem[Chaplin {\it et~al.\/}(2001)]{Chapli2001}
\biblab{Chapli2001}
Chaplin W. J., Elsworth Y., Isaak G. R., Marchenkov K. I., 
Miller B. A. \& New R., 2001,
{\it MNRAS}, {\bf 322}, 22 

\bibitem[Chaplin {\it et~al.\/}(2005)]{Chapli2005}
\biblab{Chapli2005}
Chaplin W. J., Houdek G., Elsworth Y., Gough~D.~O., Isaak G. R. \&
New R., 2005,
{\it MNRAS}, {\bf 360}, 859 

\bibitem[Christensen-Dalsgaard \& Frandsen(1983)]{Christ1983}
\biblab{Christ1983}
Christensen-Dalsgaard J. \& Frandsen S., 1983,
{\it Solar Phys.}, {\bf 82}, 165 

\bibitem[Christensen-Dalsgaard \& Mullan(1994)]{Christ1994}
\biblab{Christ1994}
Christensen-Dalsgaard J. \& Mullan D. J., 1994,
{\it MNRAS}, {\bf 270}, 921 

\bibitem[Christensen-Dalsgaard {\it et~al.\/}(1988)]{Christ1988}
\biblab{Christ1988}
Christensen-Dalsgaard J., D\"appen W. \& Lebreton~Y., 1988,
{\it Nature}, {\bf 336}, 634 

\bibitem[Christensen-Dalsgaard {\it et~al.\/}(1996)]{Christ1996}
\biblab{Christ1996}
Christensen-Dalsgaard J., D\"appen W., Ajukov~S.~V., et al., 1996,
{\it Science}, {\bf 272}, 1286 

\bibitem[Christensen-Dalsgaard {\it et~al.\/}(2005)]{Christ2005}
\biblab{Christ2005}
Christensen-Dalsgaard J., Di Mauro M. P., Schlattl~H. \& Weiss A., 2005,
{\it MNRAS}, {\bf 356},  587 

\bibitem[Cox(1974)]{Cox1974}
\biblab{Cox1974}
Cox J. P., 1974,
{\it Rep. Prog. Phys.}, {\bf 37}, 563 

\bibitem[Daszy\'nska-Daszkiewicz {\it et~al.\/}(2002)]{Daszyn2002}
\biblab{Daszyn2002}
Daszy\'nska-Daszkiewicz J., Dziembowski W. A., Pamyatnykh A. A. \& 
Goupil M.-J., 2002,
{\it A\&A}, {\bf 392}, 151 

\bibitem[Daszy\'nska-Daszkiewicz {\it et~al.\/}(2005)]{Daszyn2005}
\biblab{Daszyn2005}
Daszy\'nska-Daszkiewicz J., Dziembowski W. A., Pamyatnykh A. A., 
Breger M., Zima W. \& Houdek~G., 2005,
{\it A\&A}, {\bf 438}, 653 

\bibitem[Delache \& Scherrer(1983)]{Delach1983}
\biblab{Delach1983}
Delache P. \& Scherrer P. H., 1983,
{\it Nature}, 
{\bf 306}, 651 

\bibitem[Deupree(1990)]{Deupre1990}
\biblab{Deupre1990}
Deupree R. G., 1990,
{\it ApJ}, {\bf 357}, 175 

\bibitem[Drake \& Testa(2005)]{Drake2005}
\biblab{Drake2005}
Drake J. J. \& Testa P., 2005,
{\it Nature}, {\bf 436}, 525 

\bibitem[Dupret(2001)]{Dupret2001}
\biblab{Dupret2001}
Dupret M. A., 2001,
{\it A\&A}, {\bf 366}, 166 

\bibitem[Dupret {\it et~al.\/}(2005)]{Dupret2005}
\biblab{Dupret2005}
Dupret M.-A., Grigahc\`ene A., Garrido R., Gabriel~M. \& Scuflaire R.,
2005,
{\it A\&A}, {\bf 435}, 927 

\bibitem[Dziembowski {\it et~al.\/}(2001)]{Dziemb2001}
\biblab{Dziemb2001}
Dziembowski W. A., Gough D. O., Houdek G. \& Sienkiewicz R., 2001,
{\it MNRAS}, {\bf 328}, 601 

\bibitem[Eggenberger {\it et~al.\/}(2005)]{Eggenb2005}
\biblab{Eggenb2005}
Eggenberger P., Maeder A. \& Meynet G., 2005,
{\it A\&A}, {\bf 440}, L9 

\bibitem[Fletcher {\it et~al.\/}(2006)]{Fletch2006}
\biblab{Fletch2006}
Fletcher S. T., Chaplin W. J., Elsworth Y., Schou J. \& Buzasi D., 2006,
{\it MNRAS}, {\bf 371}, 935 

\bibitem[Fontaine {\it et~al.\/}(2003)]{Fontai2003}
\biblab{Fontai2003}
Fontaine G., Brassard P., Charpinet S., Green~E.~M., Chayer P.,
Bill\`eres M. \& Randall~S.~K., 2003,
{\it ApJ}, {\bf 597}, 518 

\bibitem[Fontaine {\it et~al.\/}(2006)]{Fontai2006}
\biblab{Fontai2006}
Fontaine G., Brassard P., Charpinet S. \& Chayer P., 2006,
{\it Mem. Soc. Astron. Ital.}, {\bf 77}, 49 

\bibitem[Fr\"ohlich {\it et~al.\/}(1984)]{Frohli1984}
\biblab{Frohli1984}
Fr\"ohlich C. \& Delache P. 1984,
{\it Mem. Soc. Astron. Ital.}, {\bf 55}, 99 

\bibitem[Gabriel {\it et~al.\/}(1995)]{Gabrie1995}
\biblab{Gabrie1995}
Gabriel A. H., Grec G., Charra J., et al., 1995,
{\it Solar Phys.}, {\bf 162}, 61 

\bibitem[Georgobiani {\it et~al.\/}(2003)]{Georgo2003}
\biblab{Georgo2003}
Georgobiani D., Stein R. F. \& Nordlund {\AA}., 2003,
{\it ApJ}, {\bf 596}, 698 

\bibitem[Gough(1977)]{Gough1977}
\biblab{Gough1977}
Gough D. O., 1977,
{\it ApJ}, {\bf 214}, 196 

\bibitem[Gough(1985)]{Gough1985}
\biblab{Gough1985}
Gough D. O., 1985,
in {\it Future missions in solar, heliospheric and space plasma physics},
eds E. Rolfe \& B. Battrick, ESA SP-235, p. 183 

\bibitem[Gough \& McIntyre(1998)]{Gough1998}
\biblab{Gough1998}
Gough D. O. \& McIntyre M. E., 1998,
{\it Nature}, {\bf 394}, 755 

\bibitem[Grevesse {\it et~al.\/}(2005)]{Greves2005}
\biblab{Greves2005}
Grevesse N., Asplund M. \& Sauval A. J., 2005,
in {\it Element stratification in stars: 40 years of atomic diffusion},
eds G. Alecian, O. Richard \& S. Vaclair,
{\it EAS Publ. Ser.}, {\bf 17}, 21 

\bibitem[Guenther(1991)]{Guenth1991}
\biblab{Guenth1991}
Guenther D. B., 1991,
{\it ApJ}, {\bf 375}, 352 

\bibitem[Guenther {\it et~al.\/}(2005)]{Guenth2005}
\biblab{Guenth2005}
Guenther D. B., Kallinger T., Reegen P., et al., 2005,
{\it ApJ}, {\bf 635}, 547 

\bibitem[Guzik {\it et~al.\/}(2000)]{Guzik2000}
\biblab{Guzik2000}
Guzik J. A., Kaye A. B., Bradley P. A., Cox A. N. \&
Neuforge C., 2000,
{\it ApJ}, {\bf 542}, L57 

\bibitem[Harvey(1988)]{Harvey1988}
\biblab{Harvey1988}
Harvey J. W., 1988,
in {\it Proc. IAU Symposium No~123, Advances in helio- and asteroseismology},
eds J. Christensen-Dalsgaard \& S. Frandsen,
Reidel, Dordrecht, p. 497 

\bibitem[Houdek(2000)]{Houdek2000}
\biblab{Houdek2000}
Houdek G., 2000,
in {\it Delta Scuti and related stars},
eds M. Breger \& M. H. Montgomery,
ASP Conf. Ser., {\bf 210}, p. 454 

\bibitem[Houdek \& Gough(2002)]{Houdek2002}
\biblab{Houdek2002}
Houdek G. \& Gough D. O., 2002,
{\it MNRAS}, {\bf 336}, L65 

\bibitem[Howard \& LaBonte(1980)]{Howard1980}
\biblab{Howard1980}
Howard R. \& LaBonte B. J., 1980,
{\it ApJ}, {\bf 239}, L33 

\bibitem[Howe {\it et~al.\/}(2002)]{Howe2002}
\biblab{Howe2002}
Howe R., Komm R. W. \& Hill F., 2002,
{\it ApJ}, {\bf 580}, 1172 

\bibitem[Howe {\it et~al.\/}(2006)]{Howe2006}
\biblab{Howe2006}
Howe R., Komm R., Hill F., Ulrich R., Haber~D.~A., Hindman B. W.,
Schou J. \& Thompson~M.~J., 2006,
{\it Solar Phys.}, {\bf 235}, 1 

\bibitem[Jackson {\it et~al.\/}(2005)]{Jackso2005}
\biblab{Jackso2005}
Jackson S., MacGregor K. B. \& Skumanich A., 2005,
{\it ApJS}, {\bf 156}, 245 

\bibitem[Jim\'enez-Reyes {\it et~al.\/}(1998)]{Jimene1998}
\biblab{Jimene1998}
Jim\'enez-Reyes S. J., R\'egulo C., Pall\'e P. L. \& Roca Cort\'es T.,
1998,
{\it A\&A}, {\bf 329}, 1119 

\bibitem[Kiss {\it et~al.\/}(2006)]{Kiss2006}
\biblab{Kiss2006}
Kiss L. L., Szabo G. M. \& Bedding T. R., 2006,
{\it MNRAS}, in the press {\tt [astro-ph/0608438]}

\bibitem[Kumar {\it et~al.\/}(1996)]{Kumar1996}
\biblab{Kumar1996}
Kumar P., Quataert E. J. \& Bahcall J. N., 1996,
{\it ApJ}, {\bf 458}, L83 

\bibitem[Libbrecht(1992)]{Libbre1992}
\biblab{Libbre1992}
Libbrecht K. G., 1992,
{\it ApJ}, {\bf 387}, 712 

\bibitem[Ligni\`eres {\it et~al.\/}(2006)]{Lignie2006}
\biblab{Lignie2006}
Ligni\`eres F., Rieutord M. \& Reese D., 2006,
{\it A\&A}, {\bf 455}, 607 

\bibitem[Lin \& D\"appen(2005)]{Lin2005}
\biblab{Lin2005}
Lin C.-H. \& D\"appen W., 2005,
{\it ApJ}, {\bf 623}, 556 

\bibitem[Lindsey \& Braun(2005)]{Lindse2005}
\biblab{Lindse2005}
Lindsey C. \& Braun D. C., 2005,
{\it ApJ}, {\bf 620}, 1107 

\bibitem[Marsch {\it et~al.\/}(2005)]{Marsch2005}
\biblab{Marsch2005}
Marsch E., Marsden R., Harrison R., Wimmer-Schweingruber R. \&
Fleck B., 2005,
{\it Adv. Space Res.}, {\bf 36}, 1360 

\bibitem[Mazumdar {\it et~al.\/}(2006)]{Mazumd2006}
\biblab{Mazumd2006}
Mazumdar A., Briquet M., Desmet M. \& Aerts C., 2006,
{\it A\&A}, in the press {\tt astro-ph/0607261 v1}

\bibitem[Monteiro \& Thompson(1998)]{Montei1998}
\biblab{Montei1998}
Monteiro M. J. P. F. G. \& Thompson M. J., 1998,
in {\it Proc. IAU Symp. 185: New eyes to see inside the Sun and stars},
eds F.-L. Deubner, J.~Christensen-Dalsgaard \& D. W. Kurtz,
Kluwer, Dordrecht, p. 317 

\bibitem[Monteiro \& Thompson(2005)]{Montei2005}
\biblab{Montei2005}
Monteiro M. J. P. F. G. \& Thompson M. J., 2005,
{\it MNRAS}, {\bf 361}, 1187 

\bibitem[Nowakowski(2005)]{Nowako2005}
\biblab{Nowako2005}
Nowakowski R. M., 2005,
{\it Acta Astron.}, {\bf 55}, 1 

\bibitem[Osaki(1975)]{Osaki1975}
\biblab{Osaki1975}
Osaki Y., 1975,
{\it Publ. Astron. Soc. Japan}, {\bf 27}, 237 

\bibitem[Ot\'{\i} Floranes {\it et~al.\/}(2005)]{Oti2005}
\biblab{Oti2005}
Ot\'{\i} Floranes H., Christensen-Dalsgaard J. \& Thompson M. J., 2005,
{\it MNRAS}, {\bf 356}, 671 

\bibitem[Pamyatnykh {\it et~al.\/}(2004)]{Pamyat2004}
\biblab{Pamyat2004}
Pamyatnykh A. A., Handler G. \& Dziembowski~W.~A., 2004,
{\it MNRAS}, {\bf 350}, 1022 

\bibitem[Rabello-Soares {\it et~al.\/}(2000)]{Rabell2000}
\biblab{Rabell2000}
Rabello-Soares M. C., Basu S., Christensen-Dalsgaard J. \&
Di Mauro M. P., 2000,
{\it Solar Phys.}, {\bf 193}, 345 

\bibitem[Reese {\it et~al.\/}(2006)]{Reese2006}
\biblab{Reese2006}
Reese D., Ligni\`eres F. \& Rieutord M., 2006,
{\it A\&A}, {\bf 455}, 621 

\bibitem[Robinson {\it et~al.\/}(2003)]{Robins2003}
\biblab{Robins2003}
Robinson F. J., Demarque P., Li L. H., Kim Y.-C., Chan K. L. \&
Guenther D. B., 2003,
{\it MNRAS}, {\bf 340}, 923 

\bibitem[Rosenthal {\it et~al.\/}(1999)]{Rosent1999}
\biblab{Rosent1999}
Rosenthal C. S., Christensen-Dalsgaard J., Nordlund {\AA}.,
Stein R. F. \& Trampedach R., 1999,
{\it A\&A}, {\bf 351}, 689 

\bibitem[Roxburgh(2004)]{Roxbur2004}
\biblab{Roxbur2004}
Roxburgh I. W., 2004,
{\it A\&A}, {\bf 428}, 171 

\bibitem[Roxburgh(2005)]{Roxbur2005}
\biblab{Roxbur2005}
Roxburgh I. W., 2005,
{\it A\&A}, {\bf 434}, 665 

\bibitem[Roxburgh \& Vorontsov(2003)]{Roxbur2003}
\biblab{Roxbur2003}
Roxburgh I. W. \& Vorontsov S. V., 2003,
{\it A\&A}, {\bf 411}, 215 

\bibitem[Salpeter(1954)]{Salpet1954}
\biblab{Salpet1954}
Salpeter E. E., 1954,
{\it Austr. J. Phys.}, {\bf 7}, 373 

\bibitem[Schatzman(1993)]{Schatz1993}
\biblab{Schatz1993}
Schatzman E., 1993,
{\it A\&A}, {\bf 279}, 431 

\bibitem[Scherrer {\it et~al.\/}(1979)]{Scherr1979}
\biblab{Scherr1979}
Scherrer P. H., Wilcox J. M., Kotov V. A.,
Severny~A.~B. \& Tsap T. T., 1979,
{\it Nature}, {\bf 277}, 635 

\bibitem[Schmelz {\it et~al.\/}(2005)]{Schmel2005}
\biblab{Schmel2005}
Schmelz J. T., Nasraoui K., Roames J. K., Lippner~L.~A. \& Garst J. W.,
2005,
{\it ApJ}, {\bf 634}, L197 

\bibitem[Schou(1992)]{Schou1992}
\biblab{Schou1992}
Schou J., 1992,
{\it On the analysis of helioseismic data},
PhD Dissertation, Aarhus University.

\bibitem[Schou(1999)]{Schou1999}
\biblab{Schou1999}
Schou J., 1999,
{\it ApJ}, {\bf 523}, L181 

\bibitem[Schunker \& Cally(2006)]{Schunk2006}
\biblab{Schunk2006}
Schunker H. \& Cally P. S., 2006,
{\it MNRAS}, in the press

\bibitem[Scuflaire(1974)]{Scufla1974}
\biblab{Scufla1974}
Scuflaire R., 1974,
{\it A\&A}, {\bf 36}, 107 

\bibitem[Severny {\it et~al.\/}(1976)]{Severn1976}
\biblab{Severn1976}
Severny A. B., Kotov V. A. \& Tsap T. T., 1976,
{\it Nature}, {\bf 259}, 87 

\bibitem[Shaviv(2004)]{Shaviv2004}
\biblab{Shaviv2004}
Shaviv G., 2004,
{\it A\&A}, {\bf 418}, 801 

\bibitem[Shaviv \& Shaviv(2000)]{Shaviv2000}
\biblab{Shaviv2000}
Shaviv G. \& Shaviv N. J., 2000,
{\it ApJ}, {\bf 529}, 1054 

\bibitem[Shaviv \& Shaviv(1996)]{Shaviv1996}
\biblab{Shaviv1996}
Shaviv N. J. \& Shaviv G., 1996,
{\it ApJ}, {\bf 468}, 433 

\bibitem[Soufi {\it et~al.\/}(1998)]{Soufi1998}
\biblab{Soufi1998}
Soufi F., Goupil M. J. \& Dziembowski W. A., 1998,
{\it A\&A}, {\bf 334}, 911 

\bibitem[Spruit(2002)]{Spruit2002}
\biblab{Spruit2002}
Spruit H. C., 2002,
{\it A\&A}, {\bf 381}, 923 

\bibitem[Stein \& Nordlund(2001)]{Stein2001}
\biblab{Stein2001}
Stein R. F. \& Nordlund {\AA}., 2001,
{\it ApJ}, {\bf 546}, 585 

\bibitem[Stello {\it et~al.\/}(2006)]{Stello2006}
\biblab{Stello2006}
Stello D., Kjeldsen H., Bedding T. R. \& Buzasi D., 2006,
{\it A\&A}, {\bf 448}, 709 

\bibitem[Straka {\it et~al.\/}(2006)]{Straka2006}
\biblab{Straka2006}
Straka C. W., Demarque P., Guenther D. B., Li L. \& Robinson F. J., 2006,
{\it ApJ}, {\bf 636}, 1078 

\bibitem[Takata(2005)]{Takata2005}
\biblab{Takata2005}
Takata M., 2005,
{\it Publ. Astron. Soc. Japan}, {\bf 57}, 375 

\bibitem[Tripathy {\it et~al.\/}(2001)]{Tripat2001}
\biblab{Tripat2001}
Tripathy S. C., Kumar B., Jain K. \& Bhatnagar~A., 2001,
{\it Solar Phys.}, {\bf 200}, 3 

\bibitem[Tsytovich(2000)]{Tsytov2000}
\biblab{Tsytov2000}
Tsytovich V. N., 2000,
{\it A\&A}, {\bf 356}, L57 

\bibitem[Turck-Chi\`eze {\it et~al.\/}(2001)]{Turck2001}
\biblab{Turck2001}
Turck-Chi\`eze S., Couvidat S., Kosovichev A. G., et al., 2001,
{\it ApJ}, {\bf 555}, L69 

\bibitem[Turck-Chi\`eze {\it et~al.\/}(2004)]{Turck2004}
\biblab{Turck2004}
Turck-Chi\`eze S., Garc\'{\i}a R. A., Couvidat S., et al., 2004,
{\it ApJ}, {\bf 604}, 455 
[Erratum: {\it ApJ}, {\bf 608}, 610]

\bibitem[Vorontsov \& Jefferies(2005)]{Voront2005}
\biblab{Voront2005}
Vorontsov S. V. \& Jefferies S. M., 2005,
{\it ApJ}, {\bf 623}, 1202 

\bibitem[Walker {\it et~al.\/}(2003)]{Walker2003}
\biblab{Walker2003}
Walker G., Matthews J., Kuschnig R., et al., 2003,
{\it Publ. Astron. Soc. Pacific}, {\bf 115}, 1023 

\bibitem[Zahn(1992)]{Zahn1992}
\biblab{Zahn1992}
Zahn J.-P., 1992,
{\it A\&A}, {\bf 265}, 115 

\end{thebibliography}

\raggedright

\end{document}